\documentclass[english,a4paper]{article}

\usepackage[a4paper]{geometry}
\usepackage{cite}
\usepackage{babel}
\usepackage{url}
\usepackage{amsmath}
\usepackage{amsbsy}
\usepackage{graphicx}
\usepackage{setspace}
\usepackage[unicode=true,
 bookmarks=true,bookmarksnumbered=false,bookmarksopen=false,
 breaklinks=false,pdfborder={0 0 1},backref=false,colorlinks=false]
 {hyperref}

\newcommand{\rmd}{\mathrm{d}}
\usepackage[dvipsnames]{xcolor}

\title{Spurious memory in non-equilibrium stochastic models of imitative behavior}
\author{Vygintas Gontis, Aleksejus Kononovicius\\Institute of Theoretical Physics and Astronomy, Vilnius University}
\date{}

\begin{document}

\maketitle

\begin{abstract}
The origin of the long-range memory in the non-equilibrium systems is still an open problem as the phenomenon can be reproduced using models based on Markov processes. In these cases a notion of spurious memory is introduced. A good example of Markov processes with spurious memory is stochastic process driven by a non-linear stochastic differential equation (SDE). This example is at odds with models built using fractional Brownian motion (fBm). We analyze differences between these two cases seeking to establish possible empirical tests of the origin of the observed long-range memory. We investigate probability density functions (PDFs) of burst and inter-burst duration in numerically obtained time series and compare with the results of fBm. Our analysis confirms that the characteristic feature of the processes described by a one-dimensional SDE is the power-law exponent $3/2$ of the burst or inter-burst duration PDF. This property of stochastic processes might be used to detect spurious memory in various non-equilibrium systems, where observed macroscopic behavior can be derived from the imitative interactions of agents.
\end{abstract}

\section{Introduction}

The application of statistical physics to diverse
fields such as social sciences and economics, biology and population genetics, medicine, information technology, computer science and others \cite{Ausloos1999PhysA,Blythe2007JStatMech,Nishimori2008Oxford,Castellano2009RevModPhys} is making this interdisciplinary research very universal. Though the number of agents is usually incomparable with number of particles in physical systems, the understanding of macroscopic behavior of social and biological systems naturally invokes methods of statistical physics with very simplified interactions of individuals. Humans and biological entities are itself complex systems with unknown detailed behavior and any attempt to reproduce microscopic interactions of agents might appear to be unrealistic. Thus the probabilistic description of agent interactions in social systems seems to be the most appropriate and natural. Even in very simplified models of agent interactions the collective behavior may lead to the ordered or disordered states. How the disordered interactions of agents create order in macroscopic behavior is very interesting question, nevertheless, the cases when non-equilibrium fluctuations do not disappear in the system are of high importance as well \cite{Alfarano2009Dyncon}.

There is a limited number of solvable and mathematically tractable models of many-body systems. Ising model is among the most fundamental and well known examples of such models \cite{Onsager1944PhysRev,Glauber1963JMathPhys}. Being a very popular tool for the investigation of transition from order and disorder states, Ising model in one-dimensional case with local interactions of spins can be simplified by Glauber dynamics \cite{Castellano2009RevModPhys} and gives motivation to many other applications of statistical mechanics. It is possible to simplify pairwise interactions of agents in the way, which leads to the solvable cases of many body systems in more spatial dimensions or topologies. The voter model is a good example of such social system widely used in modeling of opinion dynamics and population genetics \cite{Clifford1973Biometrica,Holley1975AnnProb,Redner2001Cambridge,Slanina2014Oxford}. In one-dimensional case the voter model coincides with one-dimensional Glauber dynamics,  while in other dimensions and various topologies can be solved as well. The standard voter model converges to the consensus of opinions and this is related with two circumstances: local nature of interactions and imitation of neighbor opinion without idiosyncratic decision making \cite{Slanina2014Oxford}. From our point of view, the case of global agent interactions or system running on the randomly generated network, including idiosyncratic switching of opinions, is of great importance as exhibits continuing stochastic fluctuations in collective behavior \cite{Alfarano2009Dyncon,Kononovicius2014EPJB}. The evolution of such system can be described using the Fokker-Planck equation or as a non-linear SDE for population evolution and can be seen as a special case of voter model \cite{Liggett1999Springer,Fernandez2014PhysRevLett}, Moran model \cite{Blythe2007JStatMech} or Kirman model \cite{Kirman1993QJE,Alfarano2009Dyncon}. The continuing fluctuations in such non-equilibrium system with imitative behavior of agents exhibit very general power-law scaling properties including spurious memory applicable to social \cite{Aoki2007Cambridge,Kononovicius2017Arxiv}, financial \cite{Alfarano2005CompEco,Gontis2014PlosOne,Gontis2016PhysA} or biological \cite{Blythe2007JStatMech} systems. 

Here we investigate the long-range memory property, which might originate from the true long-range memory process, one with correlated increments, such as the fractional Brownian motion (fBm) \cite{Mandelbrot1968SIAMR,Dieker2003fbm,McCauley2006PhysA,McCauley2007PhysA}. While the long-range memory might be also obtained from the stochastic processes with non-stationary uncorrelated increments \cite{Gontis2004PhysA,McCauley2006PhysA,McCauley2007PhysA,Ruseckas2011PRE}.
Note that both stochastic processes are one-dimensional unlike in other known cases of spurious memory \cite{Lanouar2011IJBSS}. There is a fundamental problem to find out which of the possible alternatives, fBm or diffusive processes with non-stationary increments, is the origin of the observed long-range memory. Here we employ the dependence of first passage time PDF on Hurst parameter $H$ for the fBm \cite{Ding1995PhysRevE,Metzler2014Springer} and apparently different behavior for the non-linear diffusive processes \cite{Redner2001Cambridge,Jeanblanc2009Springer,Gontis2012ACS}.
This explains that the long-range memory present in social, financial and biological systems could arise from non-linear agent interactions as well as from the implicit non-linear transformations of the latent variables \cite{Kaulakys2015MPLB}.

In Section \ref{sec:ImitativeBehavior} we present the generalized model of imitative (herding) behavior, in Section \ref{sec:BurstInerburst} we analyze the statistics of the burst and inter-burst durations  and in Section \ref{sec:conclusion} we discuss and conclude results.
 
\section{Non-equilibrium stochastic fluctuations arising from the imitative behavior of agents}
\label{sec:ImitativeBehavior}

One agent (particle) jump Markov processes have become an efficient tool in modeling of physical, biological and social systems \cite{Risken1996Springer,Blythe2007JStatMech,Aoki2007Cambridge}.  The microscopic behavior of agent is replaced by continuous time Markov processes with specified transition rates. In the system with large number of agents $N$ and two possible states (choices of opinion), e.g. $\{1,2\}$, there are two possible one step transitions: a) the number of agents $n$ in state 1 increases (birth) or b) decreases (death). Such a simple but general enough definition of opinion or population dynamics can be specified by two system-wide one step transition rates
\begin{gather}
P(n\rightarrow n+1)=(N-n) \mu_1(n,N)\equiv P^{+}(n,N), \nonumber \\
P(n\rightarrow n-1)=n \mu_2(n,N)\equiv P^{-}(n,N).
\label{eq:rates}
\end{gather}
In the above $\mu_i$ are per-agent transition rates to a state given by index $i$, which in general might take many different forms dependent on $n$ and $N$ \cite{Aoki2007Cambridge}. For the sake of notational simplicity we will drop explicit statement of this dependence and use shorthand, i.e. further in this paper $\mu_i \equiv \mu_i(n,N)$. We will postpone assigning an explicit form for the $\mu_i$ until it will be necessary.

The system-wide rates define the master equation for PDF of macroscopic state evolution $p(n,t)$
\begin{align}
\frac{\partial p(n,t)}{\partial t} = & P^{+}(n-1,N) p(n-1,t) + P^{-}(n+1,N) p(n+1,t) \nonumber \\
& - [P^{+}(n,N)+P^{-}(n,N) ]p(n,t).
\label{eq:master}
\end{align}
In the limit of large $N$ values one can introduce normalized system state variable $x=\frac{n}{N}$ and write down the following Fokker-Planck equation
\begin{equation}
\frac{\partial p(x,t)}{\partial t} = \frac{\partial}{\partial x}\left[x \mu_2-(1-x)\mu_1 \right] p(x,t) + \frac{1}{2N}\frac{\partial^2}{\partial x^2}\left[(1-x)\mu_1 + x\mu_2 \right]p(x,t),
\label{eq:FP}
\end{equation}
which corresponds to the following SDE in Ito sense:
\begin{equation}
\rmd x=\left[(1-x) \mu_1 - x \mu_2 \right] \rmd t + \sqrt{\frac{(1-x) \mu_1 + x \mu_2}{N}} \rmd W,
\label{eq:SDE}
\end{equation}
where $W$ is the standard Wiener noise. One gets very well known result, Kirman model on complete or random (Erdos-Renyi) graph, when per-agent transition rates are defined as follows \cite{Kirman1993QJE,Alfarano2009Dyncon,Kononovicius2014EPJB}
\begin{equation}
\mu_1= h (\varepsilon_1 + N x),\qquad \mu_2= h [ \varepsilon_2 + N (1-x)]. \label{eq:KirmanMu}
\end{equation}
In the above $\varepsilon_i$ describe idiosyncratic transition rates to state $i$ normalized by the imitation parameter and time scale on which  fluctuations occur, in the literature denoted as $h$. The same equations describe the behavior of the noisy voter model on a complete graph, see the discussion in \cite{Kononovicius2017Arxiv}. In this paper we refer to this model, defined by Eq.~(\ref{eq:KirmanMu}), and its later generalizations as an imitative behavior model, because model's per-agent transition rates to a given state depend on the number of agents already in that state. One could see this as agents copying the state (opinion) of the agents in the other state (of other opinion). In other papers this feature is oft-described using a different, yet synonymous, term herding behavior \cite{Alfarano2005CompEco,Carro2015,Kononovicius2014EPJB,Alfarano2009Dyncon}.

This model leads to a very general case in opinion and population dynamics retaining the perpetual fluctuations in system state variable $x$ even in the limit $N\rightarrow \infty$. This is ensured by the same form of the imitation term in both system-wide transition rates $N x (1-x)$ and its linear dependence on $N$. In this case the drift term in the corresponding SDE does not depend on imitation term, while idiosyncratic terms vanish in the diffusion term. If this dependence was sub-linear, then the model would instead converge to some fixed value of $x$ \cite{Kononovicius2014EPJB,Alfarano2009Dyncon}.

This model can be generalized introducing non-linear dependence of time scale on system state variable $x$, as was proposed in few financial market models to account for the intrinsic variability of trading activity \cite{Gontis2006JStatMech,Gontis2007PhysA,Kononovicius2012PhysA}. Such additional non-linearity of the system probably is common feature of the real world. From our point of view this might be considered as a source of spurious memory, which has to be identified from the empirical data of real social, biological or physical systems, e.g., see \cite{Gontis2014PlosOne} and other references there. Let us generalize per-agent transition rates by introducing non-linear inter-agent interaction time scale variability dependent on system state variable $x$ (non-linearity is controlled by a parameter $\alpha$),
\begin{equation}
\mu_1= h (\varepsilon_1 + N x) x^{-\alpha} (1-x)^{-\alpha}, \qquad \mu_2= h [ \varepsilon_2 + N (1-x)] x^{-\alpha} (1-x)^{-\alpha}. \label{eq:KirmanMuGen}
\end{equation}
Certainly, there are other choices how to introduce this variability, e.g. \cite{Kononovicius2012PhysA,Gontis2014PlosOne}. In this paper we have chosen this one as it is symmetric in respect to $x$. This feature allows us to consider the burst durations to be statistically equivalent to the inter-burst durations and retains symmetry regarding the new variable $y$ introduced below. In this paper we use $\alpha=2$, as it seems to be the most appropriate value in the modeling of return in financial markets \cite{Gontis2014PlosOne}, while other positive values of $\alpha$ could be also considered as appropriate for the modeling of social systems with inherent imitative behavior. The stationary distribution of $x$ in the generalized imitative behavior model is Beta distribution, with parameter values equal to $\varepsilon_1+\alpha$ and $\varepsilon_2+\alpha$,
\begin{equation}
p(x) = \frac{\Gamma(\varepsilon_1+\varepsilon_2+2\alpha)}{\Gamma(\varepsilon_1+\alpha)\Gamma(\varepsilon_2+\alpha)} x^{\varepsilon_1+\alpha-1} (1-x)^{\varepsilon_2+\alpha-1} .
\label{eq:BetaDistr}
\end{equation}

In earlier works \cite{Alfarano2005CompEco,Kononovicius2012PhysA,Gontis2014PlosOne,Gontis2016PhysA} it was shown that Kirman model is suitable to model return in financial markets, with return defined to be proportional to $y=\frac{x}{1-x}$. For this reason we are more interested in the properties of the time series of $y$ instead of $x$. Though other non-linear transformations of $x$, leading to very big fluctuations in the transformed time series, might reproduce similar properties as well.

Using Ito lemma \cite{Gardiner2009Springer} we can derive SDE for $y$, assuming that $x$ time series are generated by SDE (\ref{eq:SDE}) with per-agent transition rates $\mu_i$ given by Eq. (\ref{eq:KirmanMuGen}),
\begin{equation}
\rmd y=\left(\varepsilon_1 y^{-\alpha} + (2-\varepsilon_2)y^{1-\alpha}\right)(y+1)^{2\alpha+1} \rmd t + \sqrt{2y^{1-\alpha}}(y+1)^{\alpha+1} \rmd W,
\label{eq:SDEy}
\end{equation}
Note that being derived from SDE symmetric in respect to $x$ ($x$ and $1-x$ are statistically equivalent), this SDE for $y$ also has symmetry in respect to $y$. Namely $y$ and $\frac{1}{y}$ are equivalent in their statistical properties. This is important as we seek to retain the symmetry in burst and inter-burst duration statistics and in definition of $y$.

Note that SDE~(\ref{eq:SDEy}) does not satisfy Lipschitz condition. In order to satisfy Lipschitz condition, and to avoid over-flow problems in numerical simulations, we introduce reflective boundary condition for very large value of $y$. Exact choice of $y$ value to place boundaries does not influence the results discussed in this paper, yet, for sake of completeness, we would like to note that we have placed boundaries at $y_{min}=10^{-5}$ and $y_{max}=10^{5}$. Numerically we solve this SDE using Euler-Maruyama method, but using variable time steps. For more details see earlier works, e.g. \cite{Gontis2010PhysA,Kononovicius2012PhysA,Gontis2014PlosOne}, in which similar equations were solved numerically.

As can be seen from the previous discussion agent-based models with non-linear interactions can lead to the macroscopic description by a non-linear SDEs, which represent a class of Markov processes. Note that if we take only highest powers of $y$ in SDE~(\ref{eq:SDEy}), we would obtain an SDE of a well-known form \cite{Ruseckas2014JStatMech}
\begin{equation}
\rmd y=\left( \eta-\frac{\lambda}{2} \right) y^{2\eta-1} \rmd t +y^{\eta} \rmd W, \label{eq:SDE2}
\end{equation} 
where new parameters $\eta$ and $\lambda$ are related to the previously used parameters as
\begin{equation}
\eta= \frac{3+\alpha}{2}, \qquad \lambda= \varepsilon_2 + \alpha + 1. \label{eq:Parameters}
\end{equation}
SDEs having form similar to SDE~(\ref{eq:SDE2}) were considered in numerous earlier papers and it was shown that they generate the time series with the power-law statistical properties, namely PDF and power-spectral density (PSD), \cite{Kaulakys2005PhysRevE,Ruseckas2014JStatMech}:
\begin{equation}
p(x)\sim x^{-\lambda},\qquad S(f)\sim \frac{1}{f^\beta},\quad \beta=1+\frac{\lambda-3}{2\eta-2}=2H+1.
\label{eq:power-law}
\end{equation}

\begin{figure}[!h]
\centering
\includegraphics[width=0.9\textwidth]{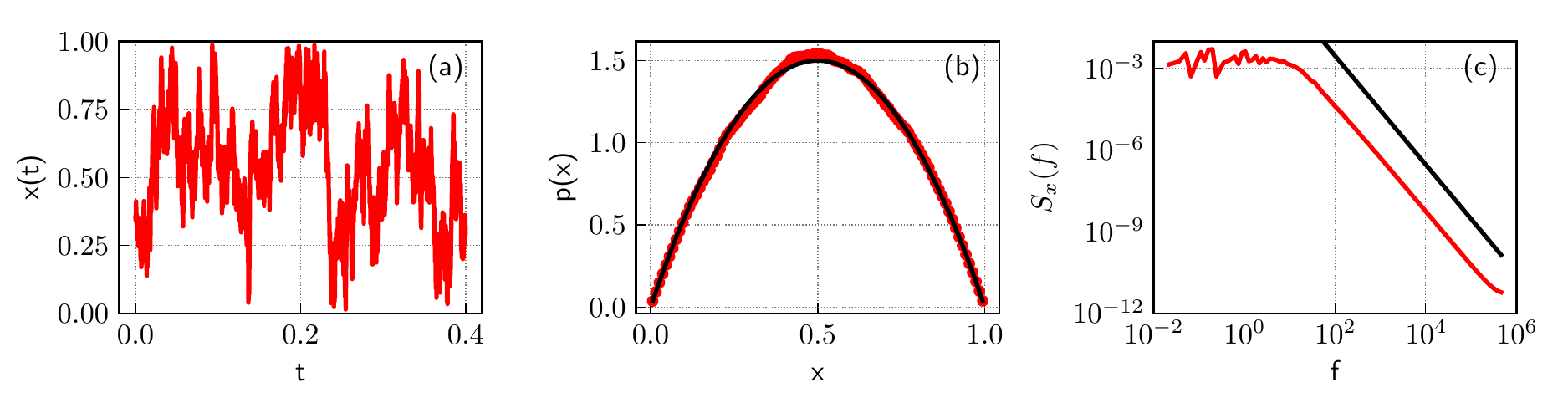}
\caption{Excerpt of time series (a), obtained by solving SDE (\ref{eq:SDE}) with $\mu_i$ given by Eq. (\ref{eq:KirmanMuGen}), (b) PDF and (c) PSD of the same time series (red curves). Black curves in (b) and (c) represent theoretical fits: (b) Beta distribution with parameter values $2$ and $2$, (c) $1/f^2$ trend line. Parameter set used in numerical simulation: $\alpha=2$, $\varepsilon_1 = \varepsilon_2 = 0$.}
\label{fig:x-alpha2}
\end{figure}

\begin{figure}[!h]
\centering
\includegraphics[width=0.9\textwidth]{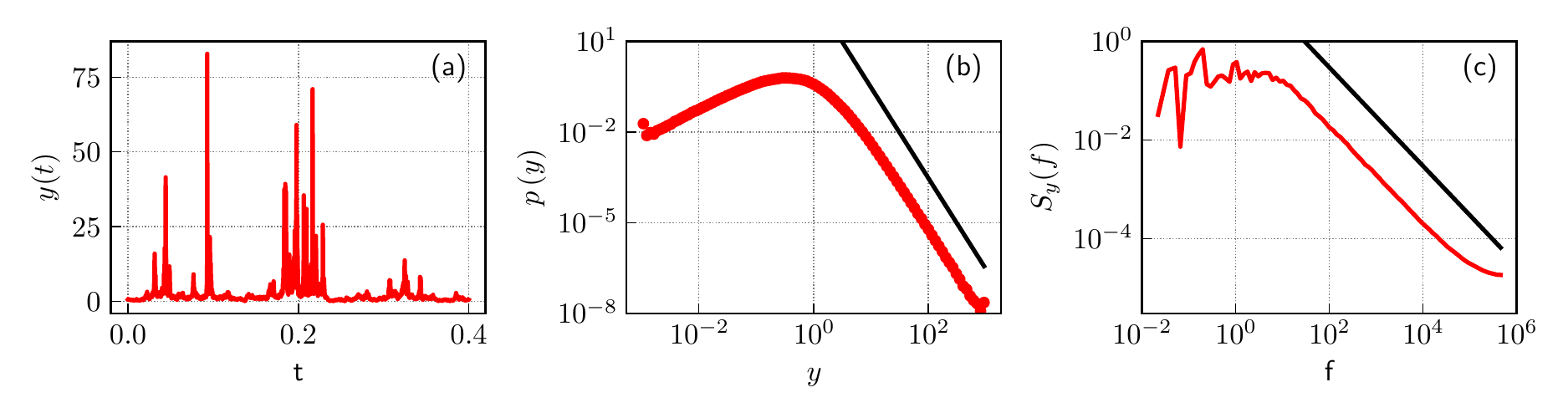}
\caption{Excerpt of the transformed $y(t) = \frac{x(t)}{1-x(t)}$ time series (a), where $x(t)$ time series is the same as in Fig. \ref{fig:x-alpha2}, (b) PDF and (c) PSD of the same transformed time series (red curves). Black curves in (b) and (c) represent theoretical fits: (b) $y^{-3}$ trend line and (c) $1/f$ trend line.}
\label{fig:y-alpha2}
\end{figure}

In Fig. \ref{fig:x-alpha2} we demonstrate excerpt of the model time series $x(t)$ (a), its stationary PDF (b) and PSD (c). Note that diffusion here is restricted in the region $0<x<1$ and PSD is $S_x\sim 1/f^2$. Only after non-linear transformation of the time series $y=\frac{x}{1-x}$, see Fig. \ref{fig:y-alpha2}, PSD becomes $S_y \sim 1/f$ and stationary PDF has power-law tail as given by Eq. (\ref{eq:power-law}). 

From our point of view spurious long-range memory might originate in many social systems with imitative behavior of agents, where long-term stochastic agent population or opinion fluctuations do not disappear. First of all such approach has to be considered as an explanation of observed long-range memory in the financial markets \cite{Gontis2010PhysA,Kononovicius2012PhysA,Gontis2016PhysA}. 
In the next section we investigate the PDF of first passage times seeking to demonstrate that such long-range memory property is different from the one observed in case of fBm. 

\section{PDFs of burst and inter-burst duration in stochastic model of imitative behavior}
\label{sec:BurstInerburst}

The class of SDEs, defined by Eq.~(\ref{eq:SDE2}), describes multifractal stochastic processes \cite{Kononovicius2012PhysA} with non-stationary increments \cite{McCauley2006PhysA,McCauley2007PhysA}, power-law PSD and related auto-correlation  (\ref{eq:power-law}) thus unlike for processes with correlated increments such as fBm, this class can be considered as having spurious memory. Here we will demonstrate clear distinction between these two different models with correlated (fBm) and  uncorrelated (non-linear SDE) increments.  
The idea of such distinction is based on the PDF of first return times
of stochastic processes $x(t)$ with absorbing boundary at some threshold level $x=h$. Ding and Yang have considered the problem for fBm as a first return time in \cite{Ding1995PhysRevE}. 

To our knowledge we make first attempt to discriminate between these two one-dimensional stochastic processes. Other models of spurious memory are qualitatively different and usually related with double stochasticity \cite{Lanouar2011IJBSS}. The exponent $3/2$ of PDF for first passage times in one-dimensional Markov processes is a characteristic feature \cite{Redner2001Cambridge,Jeanblanc2009Springer,Gontis2012ACS}, when deviations from this law is a characteristic feature of fBm \cite{Ding1995PhysRevE,Metzler2014Springer}. 
 
\begin{figure}[!h]
\centering
\includegraphics[width=0.4\textwidth]{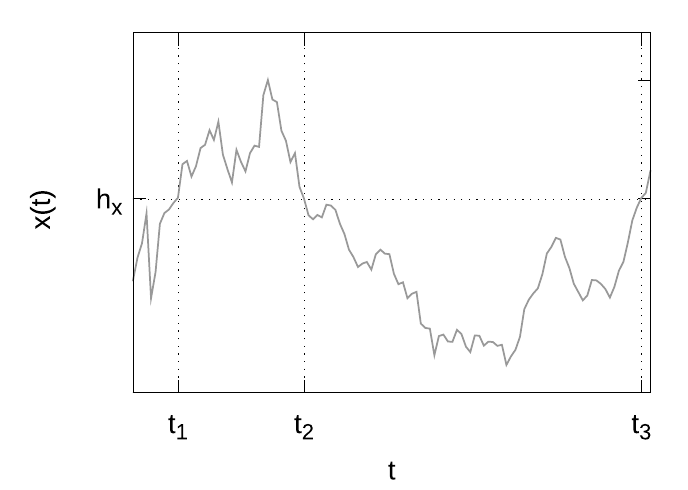}
\caption{Excerpt of a generic time series. Three threshold, $h_x$, passage events, $t_i$, are shown. Thus burst duration $T$ can be defined as $T=t_2-t_1$  and inter-burst duration $\theta$ can be defined as  $\theta=t_3-t_2$.}
\label{fig:burst-explain}
\end{figure}

In this paper we consider two distinct threshold passage events -- one describes return to the threshold from above, while the other describes return to the threshold from below. We consider burst duration as the amount of time the series spent above the threshold. In Fig. \ref{fig:burst-explain} burst period lasts from threshold passage at time $t_1$ until the next threshold passage at time $t_2$. Hence burst duration, in this case, is given by $T=t_2-t_1$. Evidently it is the same as the first return time, but with restriction that the return was made from above the threshold. In Fig. \ref{fig:burst-explain} inter-burst period lasts from threshold passage at time $t_2$ until the next threshold passage at time $t_3$. Hence inter-burst duration, in this case, is given by $\theta=t_3-t_2$. Evidently it is the same as the first return time, but with restriction that the return was made from below the threshold.

While these definitions might seem puzzling from purely theoretical point of view, they make better sense from empirical point of view. When analyzing generated, numerically or extracted from empirical data, time series each first return time, obtained after each threshold passage event, is always either burst duration or inter-burst duration. Furthermore almost surely after each burst period comes inter-burst period and vice versa. Also, as will be shown later in this paper, distribution of burst and inter-burst duration coincide (actually it also coincides with first return time distribution) if model is symmetric in respect to the selected threshold.

Due to symmetry, for all real thresholds of one-dimensional fBm its PDFs of burst and inter-burst duration coincide and can be written as \cite{Ding1995PhysRevE,Metzler2014Springer} 
\begin{equation}
p(T) \sim T^{H-2}.
\label{eq:TPDF}
\end{equation}
The Hurst parameter $H$, defining the exponent of power-law PDF $H-2$, coincides with the corresponding exponent for other one-dimensional Markov processes only when $H=\frac{1}{2}$  \cite{Borodin2002Birkhauser,Jeanblanc2009Springer,Gardiner2009Springer,Redner2001Cambridge}. When fBm is biased, the PDF of $T$ has exponential cutoff \cite{Ding1995PhysRevE}
\begin{equation}
p(T) \sim T^{H-2} \exp\left(-\frac{T}{T_s}\right),
\label{eq:TPDFbiased}
\end{equation}
where $T_s$ is defined by the mean and standard deviation of increments and Hurst parameter $H$. Note that for biased fBm PDFs of burst and inter-burst duration will no longer coincide. In this paper we generate biased fBm time series by solving the following iterative equation:
\begin{equation}
x_{i+1} = x_{i} - \gamma x_i \Delta t + \xi^{(H)}_{i} \left(\Delta t \right)^H. \label{eq:fbm-gen}
\end{equation}
In the above $\gamma$ is damping coefficient, while $\xi^{(H)}_{i}$ is fractional Gaussian noise with Hurst exponent $H$. Fractional Gaussian noise, in this case, is generated by using approximate circulant method introduced in \cite{Dieker2003fbm}.

Analytically only burst duration of SDE~(\ref{eq:SDE2}) was considered in previous work \cite{Gontis2012ACS}. Namely in \cite{Gontis2012ACS} we have shown that SDE~(\ref{eq:SDE2}) can be transformed into Bessel process. This was used to obtain the approximation of burst duration PDF, as some formulas for the distribution of first passage time of Bessel process are well known. The asymptotic behavior of the obtained burst duration PDF is given by
\begin{gather}
p_{h_x}^{(\nu)}(T) \sim T^{-3/2}, \quad \text{for} \quad 0 < T \ll \frac{2}{(\eta-1)^2 h_x^{2(\eta-1)}j_{\nu,1}^2},\label{eq:burst} \\
p_{h_x}^{(\nu)}(T) \sim \frac{1}{T}\exp\left(-\frac{(\eta-1)^2 h_x^{2(\eta-1)}j_{\nu,1}^2 T}{2}\right) , \quad \text{for} \quad T \gg \frac{2}{(\eta-1)^2 h_x^{2(\eta-1)}j_{\nu,1}^2}.
\label{eq:burst2}
\end{gather}
Here, $\nu=\frac{\lambda-2\eta+1}{2(\eta-1)}$, and $j_{\nu,1}$ is a first zero of a Bessel function of the first kind. The power-law behavior with exponent $3/2$ in Eq. (\ref{eq:burst}) is consistent with the general theory of the first passage times in one-dimensional stochastic processes \cite{Redner2001Cambridge,Jeanblanc2009Springer}.

As we have discussed in earlier sections the PDF and PSD of the time series change significantly with non-linear transformations of system state variables. Yet the distribution of first return time remains invariant under the considered transformations, as long threshold is also appropriately transformed. This may serve us as a criteria how to identify from empirical data which model of the time series is better suited to describe real system exhibiting observed long-range memory properties, as any deviation of the PDF exponent from $3/2$ other than cutoff in the region of extremely long duration would be a proof that true long-range memory is present. Thus we investigate here statistical properties of time series generated by non-linear transformations of SDE (\ref{eq:SDE}) with transition rates Eq. (\ref{eq:KirmanMuGen}), which give SDE (\ref{eq:SDEy}).

\begin{figure}[!h]
\centering
\includegraphics[width=0.6\textwidth]{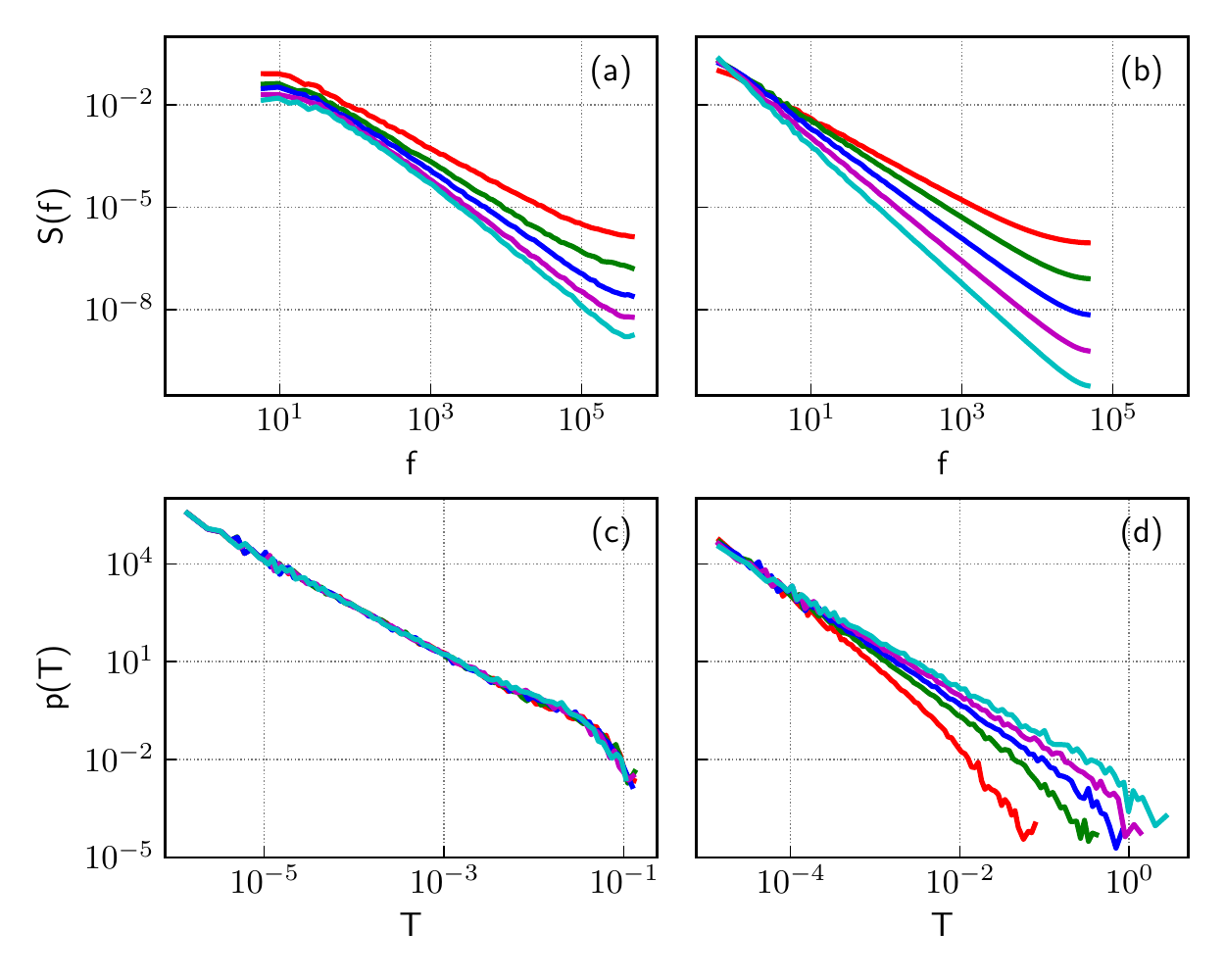}
\caption{Comparison of PSD ((a) and (b)) and burst duration PDF ((c) and (d)) of time series generated by numerically solving Eq. (\ref{eq:SDEy}) ((a) and (c)) with the ones obtained from fBm with relaxation time series, obtained by solving Eq. (\ref{eq:fbm-gen}), ((b) and (d)). fBm parameter sets: $\gamma=2$ (all cases), $H=0.1$ (red curve), $0.2$ (green curve), $0.3$ (blue curve), $0.4$ (magenta curve), $0.5$ (cyan curve). Parameters of Eq. (\ref{eq:SDEy}) are selected to give values of $H=(\beta-1)/2$ the same as for fBm: $\alpha=2$, $\varepsilon=0.6$ (red curve), $1.2$ (green curve), $1.8$ (blue curve), $2.4$ (magenta curve), $3$ (cyan curve)}
\label{fig:fBM}
\end{figure}

We demonstrate in Fig. \ref{fig:fBM} this clear distinction of fBm and SDE by numerical comparison of signal PSD and PDF of $T$. As can be seen in sub-figures (a) and (b) we have selected parameters of both models so that we obtain similar PSDs (with the same exponents $\beta$). Yet PDFs of the burst duration are different for the considered models (sub-figures (c) and (d)). While burst duration PDFs are power-law in both cases, the exponent differ. For the case of SDE (\ref{eq:SDEy}) we have same exponent for all cases, constant at $3/2$, as predicted by Eq.~(\ref{eq:burst}). While for fBm case we have differing exponents, which coincide with predictions, $2-H$, by Eq.~(\ref{eq:TPDF}). Exponential cut-offs are present in both cases, as models are mean reverting. Thus the statistical analysis of burst and inter-burst duration in empirical time series could reveal whether empirical time series contain true long-range memory or spurious memory, which originates from non-linear Markov stochastic processes.
 
It is worth to define more precisely statistical properties of burst and inter-burst duration in described agent system with imitative behavior as potentially recoverable in real social systems \cite{Gontis2016PhysA}. 
    
\begin{figure}[!h]
\centering
\includegraphics[width=0.9\textwidth]{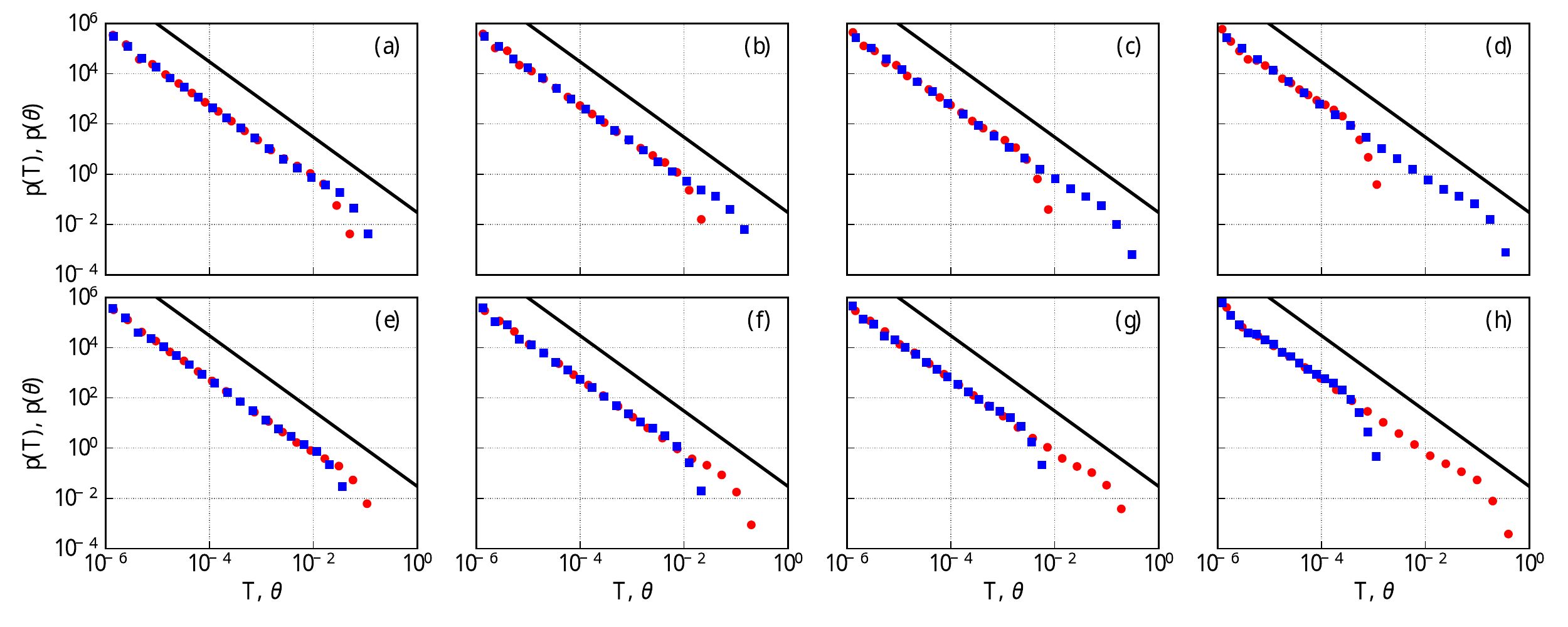}
\caption{PDFs of burst (red circles) and inter-burst (blue squares) durations of time series, obtained by solving SDE (\ref{eq:SDE}) with $\mu$ given by Eq. (\ref{eq:KirmanMuGen}), for various thresholds: $h_x=0.6$ (a), $0.7$ (b), $0.8$ (c), $0.9$ (d), $0.4$ (e), $0.3$ (f), $0.2$ (g) and $0.1$ (h). Parameter set used in numerical simulation: $\alpha=2$, $\varepsilon_1 = \varepsilon_2 = 0$. Solid black lines guide the eye according the power-law $3/2$.}
\label{fig:durationPDF}
\end{figure}

In Fig. \ref{fig:durationPDF} we present numerical calculations of PDF for burst $T$ and inter-burst $\theta$ durations with different values of the threshold $h_x$. Our numerical simulations confirm the symmetry of the model for the statistics of burst and inter-burst durations regarding values of threshold $h_x$ on both sides of mid-point $h_x=0.5$, where PDFs of $T$ and $\theta$ coincide. Note that fundamental power-law $3/2$ is retained for all values of threshold but burst and inter-burst durations have different positions of power-law cutoff. More precise analyses confirms that cutoff of PDF for burst duration $T$ is well described by previously derived exponential form Eq. (\ref{eq:TPDF}). The cutoff of inter-burst duration has the similar exponential form, but location is shifted proportionally to the $h_x$ deviation from the mid point.

Observed power-law properties of this model of imitative  behavior arise from the power-law properties of non-linear SDEs Eq. (\ref{eq:SDE2}) extensively studied elsewhere \cite{Ruseckas2010PhysRevE,Ruseckas2011PRE,Ruseckas2014JStatMech}. Note that these properties are in close relation with rapidly developing ideas of non-extensive statistical mechanics and generalized concept of entropy \cite{Ausloos2003PRE,Tsallis2009Springer}. 

\section{Conclusions}
\label{sec:conclusion}
It is widely accepted that fluctuations of volatility and trading activity in the financial markets exhibit slowly decaying auto-correlations and so-called $1/f$ noise  \cite{Engle2001QF,Plerou2001QF,Gabaix2003Nature,Ding2003Springer}. The discussion whether this slow decay corresponds to long-range memory is still ongoing. The statistical analysis in general is not able to provide a definite answer concerning the presence or absence of long-range memory in finance \cite{Lo1991Econometrica,Willinger1999FinStoch,Mikosch2003}. From our point of view, the heterogeneous agent dynamics has to be employed  seeking to explain statistical properties of financial time series \cite{Kononovicius2013EPL,Gontis2014PlosOne,Gontis2016PhysA}. Certainly, such complex system as finance \cite{Gontis2016PhysA} is not the best starting point for conceptual consideration of the long-range memory problem in other social systems. Thus in this contribution we consider much more abstract definition of agent system with imitative behavior leading to the continuing non-equilibrium  stochastic fluctuations. Derived SDEs driven by Wienner noise describe Markov processes and can't be treated as suitable to model long-range memory with correlated stochastic increments. Such modeling by stochastic agent systems becomes as an alternative to the stochastic processes driven by fBm. Thus the choice between these two possibilities is the fundamental question for understanding of the observed long-range memory property in many other complex systems.

The model we investigate here is the generalized version of Kirman model with pairwise global interaction of agents and is directly related to the voter model as well. The introduced additional feedback of macroscopic state on the time scale of interactions lets us to adjust the multiplicativity $\eta$ of SDE, defining properties of PSD for the ratio $y=\frac{x}{1-x}$. The retained symmetry of generalized equations makes this choice as preferable among other possible alternatives. From our point of view such modeling first of all is applicable to the financial systems, but is general enough and analytically tractable for other systems with heterogeneous agents.  

Here we prove analytically and numerically that PDF of  burst and inter-burst duration of stochastic variable $y$ is a power-law $3/2$ with exponential cut-off for extremely long durations. We consider this property as very valuable being very different from fBm process, where PDF of first passage time is dependent on $H$, $p(T) \sim T^{H-2}$ \cite{Ding1995PhysRevE,Metzler2014Springer}. Thus a detailed empirical analysis of burst and inter-burst duration may serve as a criteria to distinguish two different origins of $1/f$ noise and long-range memory property. Empirical evidence of the power-law exponent $3/2$ should be considered as a case of spurious memory, when deviations from this exponent should witness presence of real long-rage correlations. From our point of view the financial markets have to be considered as a social system with imitative behavior and spurious memory arising from the non-linear stochastic dynamics \cite{Gontis2014PlosOne,Gontis2016PhysA,Gontis2017PhysA}.

\end{document}